\documentclass[twocolumn,twoside,slac]{revtex4}
\usepackage{graphicx}
\usepackage{fancyhdr}
\usepackage{pifont}
\begin{document}

\title{The first deployment of workload management services on the EU
DataGrid Testbed: feedback on design and implementation.}

%

\author{G. Avellino, S. Beco, B. Cantalupo, F. Pacini, A. Terracina, A. Maraschini}
\affiliation{DATAMAT S.p.A.}

\author{D. Colling}
\affiliation{Imperial College London}

\author{S. Monforte, M. Pappalardo}
\affiliation{INFN, Sezione di Catania}

\author{L. Salconi}
\affiliation{INFN, Sezione di Pisa}

\author{F. Giacomini, E. Ronchieri}
\affiliation{INFN, CNAF}

\author{D. Kouril, A. Krenek, L. Matyska, M. Mulac, J. Pospisil, M. Ruda, Z. Salvet, J. Sitera, M. Vocu}
\affiliation{CESNET}

\author{M. Mezzadri, F. Prelz}
\affiliation{INFN, Sezione di Milano}

\author{A. Gianelle, R. Peluso, M. Sgaravatto}
\affiliation{INFN, Sezione di Padova}

\author{S. Barale, A. Guarise, A. Werbrouck}
\affiliation{INFN, Sezione di Torino}

\begin{abstract}
Application users have now been experiencing for about a year with
the standardized resource brokering services provided by the 
'workload management' package of the EU DataGrid project (WP1).
Understanding, shaping and pushing the limits of the system has provided 
valuable feedback on both its design and implementation. 
A digest of the lessons, and ``better practices", that were learned, and 
that were applied towards the second major release of the software, is given.
\end{abstract}

\maketitle

\thispagestyle{fancy}


\section{Introduction}
\begin{figure*}[t]
\centering
\includegraphics[width=135mm]{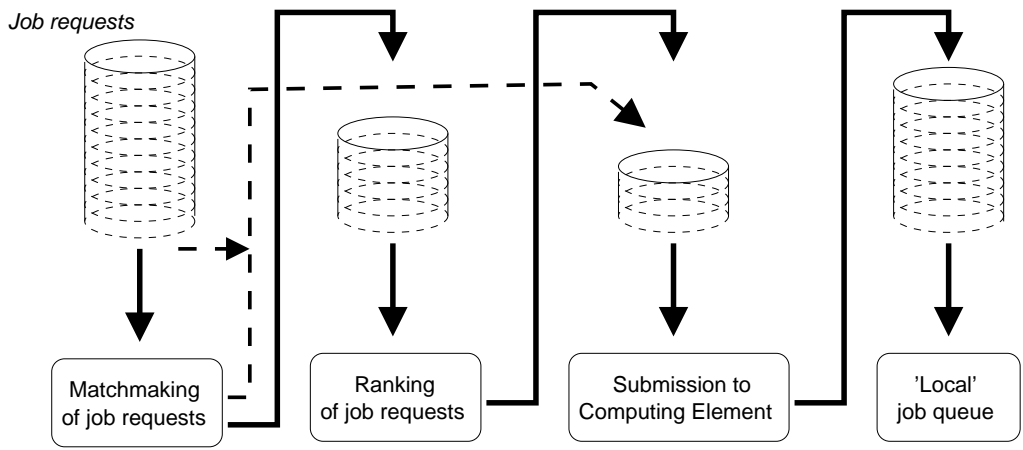}
\caption{The various steps to process a job request can be modeled
as passing the request through a network of queues.} \label{fig-netqueue}
\end{figure*}
The workload management task (Work Package 1, or WP1) \cite{wp1} of
the EU DataGrid project \cite{DataGrid} (also known, and referred to
in the following text, as EDG) is mandated to define and implement
a suitable architecture for distributed scheduling and
resource management in the Grid environment.
During the first year and a half of the project (2001-2002), and following 
a technology evaluation process, EDG WP1 defined, implemented and deployed a
set of services that integrate existing components, mostly from the
Condor \cite{Condor} and Globus \cite{Globus} projects. This was described in
more detail at CHEP 2001 \cite{WP1Chep01}. 
In a nutshell, the core job submission component of CondorG (\cite{CondorG}), 
talking to computing resources
(known in DataGrid as Computing Elements, or CEs) via the Globus GRAM
protocol, is fundamentally complemented by:
\begin {itemize}
\item A job requirement matchmaking engine (called the {\it Resource Broker},
or RB), matching job requests to computing resource status coming from the
Information System and resolving data requirements against the 
replicated file management services provided by EDG WP2.
\item A job Logging and Book-keeping service (LB), where a job state machine
is kept current based on events generated during the job lifetime, and the job
status is made available to the submitting user. The LB events
are generated with some
redundancy to cover various cases of loss.
\item A stable user API (command line, C++ and JAVA) for access to the system.
\end{itemize}
Job descriptions are expressed throughout the system 
using the Condor Classified Ad language,
where appropriate conventions were established to express requirement and
ranking conditions on Computing and
Storage Element info, and to express data requirements.
More details on the structure and evolution of these services and the
necessary integration scaffolding can be found in various EDG 
public deliverable documents.
\par
This paper focuses on how the experience of 
the first year of operation of the WP1 services on the EDG testbed was
interpreted, digested, and how a few design {\it principles} were learned 
(possibly the hard way) from the design and implementation shortcomings of 
the first release of WP1 software.
\par
These principles were applied to design and implement the second major
release of WP1 software, that is described in another CHEP 2003 paper
(\cite{WP1Chep03Sga}).
\par
To illustrate the logical path that leads to at least some of these
principles, we start by exploring the available techniques to model the 
behaviour and throughput of the integrated workload management system,
and identify two factors that significantly complicate the system analysis.

\section{The Workload Management System as a network of queues}

The workload management system provided by EDG-WP1 is designed to rely as
much as possible on existing technology. While this has the obvious
advantages of limiting effort duplication and facilitating the compatibility 
among different projects, it also significantly complicates troubleshooting
across the various layers of software supplied by different providers, and
in general the understanding of the integrated system. Also, where negotiations
with external software providers couldn't reach an agreement within the
EDG deadlines, some of the interfaces and communication paths in the system 
had to be adapted to fit the existing external software incarnations.
\par
To get a useful high-level picture of the integrated Workload Management system,
beyond all these practical constraints,
we can model it as a queuing system, where job requests traverse
a network of queues, and the ``service stations" connected to each queue
represent one of the various processing steps in the job life-cycle.
A few of these steps are exemplified in Figure \ref{fig-netqueue}.
\par
Establishing the scale factors for each service in the WP1 system 
(e.g.: how many users can a single matchmaking/job submission station serve,
how many requests per unit time can a top-level access point to the 
information system
serve, what is the sustained job throughput that can be achieved 
through the workload management chain, etc.) is one of the fundamental
premises for the correct design of the system. One could
expect to obtain this knowledge
either by applying queuing theory to this network model (this requires
obtaining a formal representation of all the components, their
service time profiles and their interconnections) or by measuring the
service times and by identifying where long queues are likely to build up
when a ``realistic" request load is injected in the system. This information
could in principle also be used to identify the areas of the system where
improvement is needed (sometimes collectively called {\it bottlenecks}).
\par
Experience with the WP1 software integration showed that both of these 
approaches are impractical for either dimensioning the system or (possibly
even more important) for identifying the trouble areas that affect the 
system throughput. We identified two non-linear factors that definitely 
work against the predictive power of queuing theory in this case, and require 
extra care even to apply straightforward reasoning when bottlenecks
are to be identified to improve system throughput.
These are the consequence
of common programming practice (and are therefore easy to be found
in the software components that we build or are integrating) and 
are described in the following Section.

\section{Troubleshooting the WMS}
\label{sec-fmode}
\begin{figure}
\includegraphics[width=80mm]{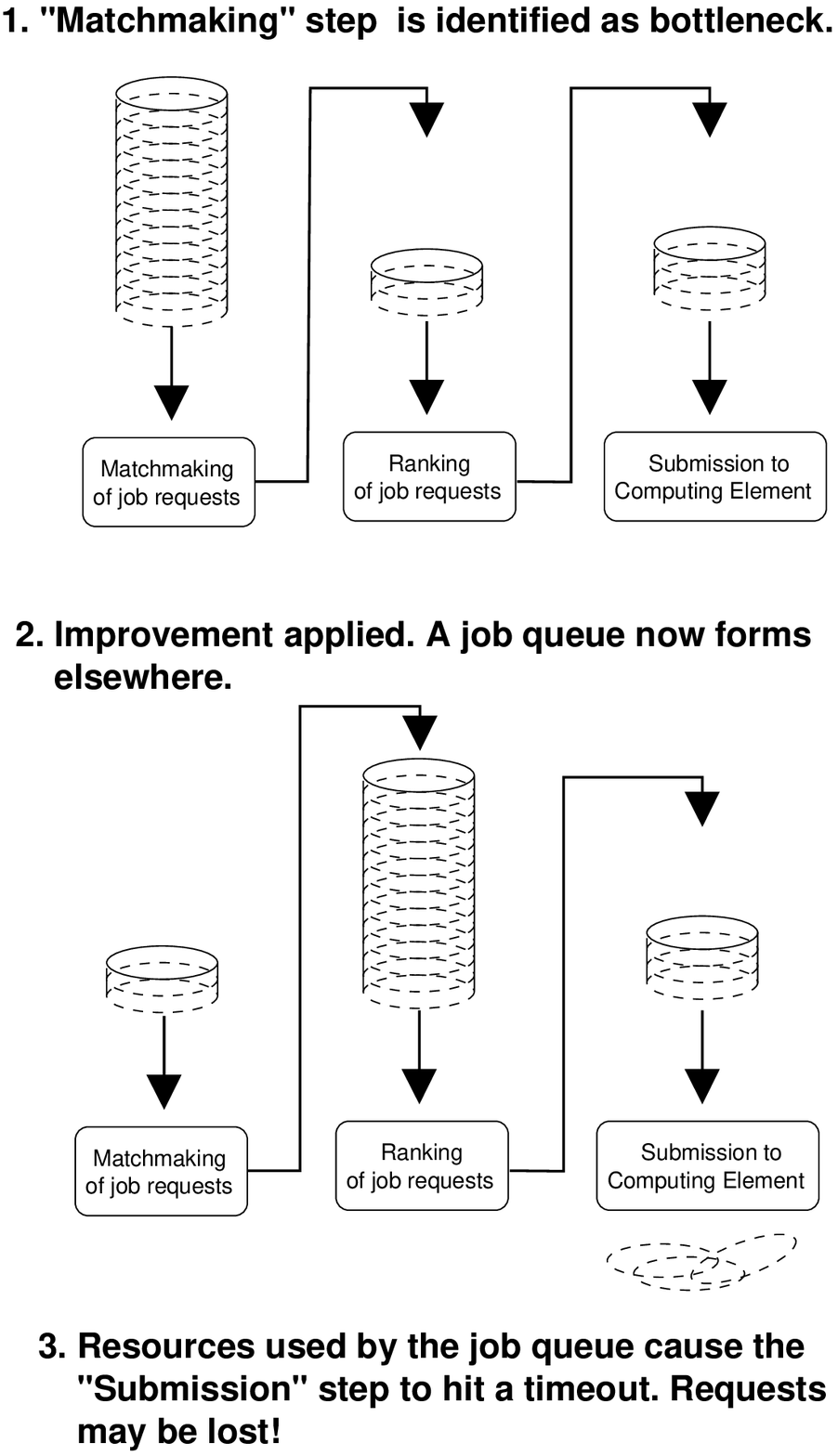}
\caption{A possible way to make the system throughput worse
by applying the genuine intent to make it better. The names
of the various steps are just an example and don't refer to 
any real experience or software component.} \label{fig-fmode}
\end{figure}
One of the most common (and most frustrating, both to developers
and to end users) experiences in troubleshooting the WP1
Workload Management system on the EDG testbed has been the fact that
often, perceived {\it improvements} to the system (sometimes even
simple bug fixes) result in a {\it decrease} in the system stability,
or reliability (fraction of requests that complete successfully).
The cause is often closely related to the known fact that removing
a bottleneck, in any flow system, can cause an overflow downstream,
possibly close to the {\it next} bottleneck. The complicating factor
is that there are at least two characteristics that could (and possibly 
still can) be found in many elements of our integrated workload management 
queuing network, that can cause problems to appear even very far from the area
of the network where an {\it improvement} is being attempted:
\begin{itemize}
\item {\it Queues of job requested can form where they can impact on the 
      system load.}\\
      Different techniques can be chosen or needed to pass 
      job requests around. Sometimes a socket connection is needed, sometimes
      sequential request processing (one request at a time in the system) 
      is required for some reason, and multiple processes/threads may be used to
      handle individual requests. Having a number of
      tasks (processes/threads) wait for a socket queue or a sequential
      processing slot is one way to ``queue" requests that definitely 
      generates much extra work for the process scheduler, and can cause 
      any other process served by the same scheduler to be allocated 
      less and less time. Queues that are unnecessarily 
      scanned while waiting for some other condition to allow the processing
      of their element can also impact on the system load, especially
      if the queue elements are associated to significant amounts of
      allocated dynamic memory.
\item {\it Some system components can enforce hard timeouts and cause
      anomalies in the job flow.}\\
      When handling the access (typically via socket connections)
      to various distributed services, provisions typically need to
      be made to handle
      all possible failure modes. ``Reasonably" long
      timeouts are sometimes chosen to handle failures that are perceived to be
      very unlikely by developers (failure to establish communication
      to a local service, for instance). This kind of failures, however,
      can easily materialise when the system resources are exhausted under
      a stress test or load peak.
\end{itemize}
Figure \ref{fig-fmode} illustrates how these two effects can conspire to
frustrate a genuine effort to remove what seems the limiting bottleneck
in the system (the example in the Figure does nor refer to any real case
or component): removing the bottleneck (1) causes a request queue to build
up at the next station (2), and this interferes via the system load to
cause hard timeouts and job failures elsewhere (3). This example
is used to rationalise some of the unexpected reactions
that, in many cases, were found while working on the WP1 integrated system.
The experience on practical troubleshooting cases similar to this one,
while bringing an understanding of the difficulties inherent in building 
distributed systems,
also drove us to formulate some of the principles that are presented 
in the next section.

\section{Principles that were learnt (and applied to improve the design)}
The attempts at getting a deeper understanding of the EDG-WP1 Workload
Management System and their failures led us to formulate a few design
principles and to apply them to the second major software release.
Here are the principles that descend from the paradigm example described in
Section \ref{sec-fmode}:
\begin{enumerate}
\renewcommand{\labelenumi}{\ding{228} \theenumi. }
\item {\bf Queues of various kinds of requests for processing should be 
allowed to form where they have a minimal and understood impact on 
system resources.}\\
Queues that get `filled' in the form of multiple threads or processes,
or that allocate significant amounts of system memory should be avoided, as they
not only adversely impact system performance, but also generate
inter-dependencies and complicate troubleshooting.
\item {\bf Limits should always be placed on dynamically allocated objects, 
threads and/or subprocesses.}\\
This is a consequence of the previous point: every dynamic resource that
gets allocated should have a tunable system-wide limit that gets enforced.
\item {\bf Special care needs to be taken around the pipeline areas where 
serial handling of requests is needed.}\\
The impact of any contention for system resources becomes more evident near
areas of the queuing system that require the acquisition of system-wide locks.
\end{enumerate}
\par
So far we concentrated on a specific attempt at modeling and understanding
the workload management system that led to an increased attention to the
usage of shared resources. There were other specific practical issues that
emerged during the deployment and troubleshooting of the system
and that led to the awareness of some fundamental design or
implementation mistake that was made. Here is a short list, where the
fundamental principle that should correct the fundamental mistake that was made
is listed:
\begin{enumerate}
\setcounter{enumi}{3}
\renewcommand{\labelenumi}{\ding{228} \theenumi. }
\item {\bf Communication among services should always be reliable:}
\begin{itemize}
  \item[-] Always applying double-commit and rollback for network communications.
  \item[-] Going through the filesystem for local communications.
\end{itemize}
In general, forms of communication that don't allow for data or messages to
be lost in a broken pipe lead to easier recovery from system or process
crashes. Where network communication is necessary, database-like techniques
have to be used.
\item {\bf Every process, object or entity related to the job lifecycle should 
have another process, object or entity in charge of its well-being.}\\
Automatic fault recovery can only happen if every entity is held accountable
and accounted for.
\item {\bf Information repositories should be minimized (with a clear 
identification of authoritative information).}\\
Many of the software components that were integrated in the EDG-WP1 solution
are stateful and include local repositories for request information, in 
the form of local queues, state files, database back-ends. Only one
site with authoritative information about requests has to be identified
and kept.
\item {\bf Monolithic, long-lived processes should be avoided.}\\
Dynamic memory programming, using languages and techniques that 
require explicit release of dynamically allocated objects, can lead to leaks
of memory, descriptors and other resources. Experimental, R\&D code
can take time to leak-proof, so it should possibly not be
linked to system components that are long-lived, as it can accelerate
system resource starvation. Short-lived, easy-to-recover components are a clean
and very practical workaround in this case.
\item {\bf More thought should be devoted to efficiently and correctly 
recovering a service rather than to starting and running it.}\\
This is again a consequence of the previous point: the capability
to quickly recover from
failures or interruption helps in assuring that system components `can' be
short-lived, either by design or by accident.
\end{enumerate}

\section{Conclusions}
EDG-WP1 has been distributing jobs over the EDG testbed in a continuous 
fashion for one and a half years now, with a software solution where
existing grid technology was integrated wherever possible.
\par
The experience of understanding the direct and indirect interplay of the service
components could not be reduced to a simple {\it scalability} evaluation.
This because understanding and removing {\it bottlenecks} is significantly 
complicated by non-linear and non-continuous effects in the system.
In this process, few principles that apply to the very complex practice
of distributed systems operations were learned the hard way (i.e. not by
just reading some good book on the subject).
EDG-WP1 tried to incorporate these principles 
in its second major software release that
will shortly face deployment in the EDG testbed.

\begin{acknowledgments}
DataGrid is a project funded by the European Commission under contract
IST-2000-25182.
\par
We also acknowledge the national funding agencies participating to
DataGrid for their support of this work.
\par
We wish to thank the Condor development team for many valuable discussions.
\end{acknowledgments}


\end{document}